 \documentclass[]{elsarticle_mod}

\usepackage{array}
\usepackage{rotating}

\usepackage{amssymb}

\begin{document}

\begin{frontmatter}



\title{Measurement of the neutrino velocity with the ICARUS detector at the CNGS beam}


\author[a]{The ICARUS Collaboration\\M.~Antonello}
\author[a]{P.~Aprili}
\author[b]{B.~Baiboussinov}
\author[b]{M.~Baldo Ceolin\footnote[1]{Deceased}}
\author[c]{P.~Benetti}
\author[c]{E.~Calligarich}
\author[a]{N.~Canci}
\author[b]{S.~Centro}
\author[e]{A.~Cesana}
\author[f]{K.~Cie\'slik}
\author[g]{D.B.~Cline}
\author[d]{A.G.~Cocco}
\author[f]{A.~Dabrowska}
\author[b]{D.~Dequal}
\author[h]{A.~Dermenev}
\author[c]{R.~Dolfini}
\author[b]{C.~Farnese}
\author[b]{A.~Fava}
\author[i]{A.~Ferrari}
\author[d]{G.~Fiorillo}
\author[b]{D.~Gibin}
\author[c]{A.~Gigli Berzolari\footnotemark[1]}
\author[h]{S.~Gninenko}
\author[b]{A.~Guglielmi}
\author[f]{M.~Haranczyk}
\author[n]{J.~Holeczek}
\author[h]{A.~Ivashkin}
\author[n]{J.~Kisiel}
\author[n]{I.~Kochanek}
\author[m]{J.~Lagoda}
\author[n]{S.~Mania}
\author[o]{G.~Mannocchi}
\author[c]{A.~Menegolli}
\author[b]{G.~Meng}
\author[c]{C.~Montanari}
\author[g]{S.~Otwinowski}
\author[o]{L.~Periale}
\author[c]{A. Piazzoli}
\author[o]{P.~Picchi}
\author[c]{F.~Pietropaolo}
\author[p]{P. Plonski}
\author[c]{A.~Rappoldi}
\author[c]{G.L.~Raselli}
\author[c]{M.~Rossella}
\author[a,i]{C.~Rubbia}
\author[e]{P.~Sala}
\author[l]{E.~Scantamburlo}
\author[e]{A.~Scaramelli}
\author[a]{E.~Segreto}
\author[q]{F.~Sergiampietri}
\author[f]{D.~Stefan}
\author[m]{J.~Stepaniak}
\author[m]{R.~Sulej}
\author[f]{M.~Szarska}
\author[e]{M.~Terrani}
\author[b]{F.~Varanini}
\author[b]{S.~Ventura}
\author[a]{C.~Vignoli}
\author[g]{H.G.~Wang}
\author[f]{A.~Zalewska}
\author[p]{K. Zaremba}
\author[i]{\\and\\P.~Alvarez~Sanchez}
\author[i]{J.~Serrano}

\address[a]{INFN - Laboratori Nazionali del Gran Sasso, Assergi (AQ), Italy}
\address[b]{Dipartimento di Fisica e INFN, Universit\`a di Padova, Via Marzolo 8, Padova, Italy}
\address[c]{Dipartimento di Fisica  e INFN, Universit\`a di Pavia, Via Bassi 6, Pavia, Italy}
\address[d]{Dipartimento di Scienze Fisiche e INFN, Universit\`a Federico II, Napoli, Italy}
\address[e]{INFN, Sezione di Milano e Politecnico, Via Celoria 16, Milano, Italy}
\address[f]{H.Niewodnicza\'nski Institute of Nuclear Physics, Krak\'ow, Poland}
\address[g]{Department of Physics and Astronomy, University of California, Los Angeles, USA}
\address[h]{Institute for Nuclear Research of the Russian Academy of Sciences, prospekt 60-letiya Oktyabrya 7a, Moscow, Russia}
\address[i]{CERN, European Laboratory for Particle Physics,  Geneve , Switzerland}
\address[l]{Universit\`a di L'Aquila, via Vetoio, Localit\`a Coppito, L'Aquila, Italy}
\address[m]{A.So\l{}tan Institute for Nuclear Studies, 05-400 Swierk/Otwock, Warszawa, Poland}
\address[n]{Institute of Physics, University of Silesia, Katowice, Poland}
\address[o]{INFN Laboratori Nazionali di Frascati, Via Fermi 40, Frascati, Italy}
\address[p]{Institute for Radioelectronics, Warsaw Univ. of Technology, Pl. Politechniki 1, Warsaw, Poland}
\address[q]{Dipartimento di Fisica e INFN, Universit\`a di Pisa, Largo Bruno Pontecorvo 3, Pisa, Italy}
\address{ }


\begin{abstract}
At the end of the 2011 run, the CERN CNGS neutrino beam  has been briefly operated in lower intensity mode with $\sim10^{12}$ p.o.t./pulse and with a proton beam structure made of four LHC-like extractions, each with a narrow width of $ \sim3$ ns, separated by 524 ns. This very tightly bunched beam  allowed a very accurate time-of-flight measurement of neutrinos from CERN to LNGS on an event-by-event basis. 
The ICARUS T600 detector (CNGS2) has collected 7 beam-associated events, consistent with the CNGS collected neutrino flux of $2.2 \times 10^{16} $ p.o.t. and in agreement with the well known characteristics of neutrino events in the LAr-TPC.  
The time of flight difference  between the speed of light and the arriving neutrino LAr-TPC events has been analysed.  The result $\delta t = 0.3 \pm 4.9 (stat.) \pm 9.0 (syst.)$ ns is compatible with the simultaneous arrival of all events with speed equal to that of light. This is in a striking difference with the reported result of OPERA~\cite{[01]} claiming that high energy neutrinos from CERN arrive at LNGS  $\sim 60$ ns earlier than expected from luminal speed. 

\end{abstract}


\end{frontmatter}


\section{Introduction}
\label{intro}
The OPERA collaboration has recently announced the measurement of the velocity of neutrinos, as they travel from CERN to the Gran Sasso Laboratory (LNGS) covering a distance of about 730 km~\cite{[01]}. OPERA (CNGS1) reported the surprising result that neutrinos arrive earlier than expected from luminal speed by a time interval  $\delta t = 57.8 \pm 7.8 (stat.)_{-5.9}^{+8.3} (syst.)$ ns. This corresponds to a superluminal propagation speed of neutrinos of an amount  $\delta c_\nu = (2.37 \pm 0.32 (stat.)_{-0.24}^{+0.34} (syst.) )\times 10^{-5}$, where $\delta c_\nu = (v_\nu - c)/c$. MINOS~\cite{[02]} had previously performed a similar measurement after a distance of 735 km and a neutrino spectrum with an average energy of 3 GeV.  Although less significant statistically, the corresponding MINOS result is $\delta c_\nu = (5.1 \pm 2.9)\times 10^{-5}$ at 68\% C.L. This measurement is consistent with the speed of light to less than 1.8$\sigma$.  The corresponding 99\% confidence interval on the speed of the neutrino is $-2.4 \times 10^{-5} \leq \delta c_\nu \leq 12.6 \times 10^{-5}$. Earlier experiments~\cite{[03]} had set upper limits of about $\delta c_\nu \leq 4 \times 10^{-5}$  in the energy range 30 to 300 GeV. 

The OPERA result has triggered new experimental determinations, since Ñ if confirmed Ñ it will call for a complete reconsideration of the basic principles of particle physics. Cohen and Glashow~\cite{[04]} have argued that such super-luminal neutrinos should lose energy by producing photons and  $e^+ e^-$ pairs, through $Z_0$ mediated processes, analogous to Cherenkov radiation. In the same CNGS neutrino beam from CERN, the ICARUS collaboration (CNGS2)~\cite{[05],[06],[07]} has recently experimentally searched for superluminal Cherenkov-like  pairs inside the fiducial volume of its large LAr-TPC detector~\cite{[08]}. No candidate event was found, setting a tight negative limit   $\delta c_\nu \leq 2.5 \times 10^{-8}$   at 90\% confidence level.  The limit for the Cohen and Glashow effect at our energies is therefore comparable to the limit on  $ \delta c_\nu$ from the lower energy SN1987A~\cite{[09],[10],[11],[12]}.

In order to initiate a new campaign of measurements on the neutrino time of flight, from October $21^{st}$ to November $4^{th}$ 2011, the CERN-SPS accelerator has been operated in a new, lower intensity mode with $\sim 10^{12}$ p.o.t./pulse and with a beam structure made of four LHC-like extractions, each with a narrow width of $ \sim$3 ns FWHM, separated by 524 ns (1/4 of the proton revolution time in the CERN-PS). This very tightly bunched beam structure represents a substantial progress since it allows a very accurate neutrino time-of-flight measurement on an event-by-event basis.

Experimental data have been collected by CNGS1 (OPERA) and CNGS2 (ICARUS) neutrino dedicated experiments, as well as by the other experiments at LNGS. We hereby report the experimental observation of the neutrino velocity distribution of the events observed in ICARUS, combining the accurate determination of the distance and time of flight with the direct observation of either neutrino events inside the detector or of neutrino associated muons from the surrounding rock.  Recently, optical triangulations have provided  the determination of the distance from OPERA origin to the ICARUS detector entry wall along the CNGS beam direction with an uncertainty of tens of centimeters. Moreover the GPS-based timing system~\cite{[01],[13]}, linking CERN and LNGS with uncertainties of a few nanoseconds, has been made available to the LNGS experiments. 

\section{Synchronizing between CERN and LNGS}
\label{syncro}
\begin{sidewaysfigure}[!htb]
\begin{center}
\includegraphics[width=0.90\textwidth]{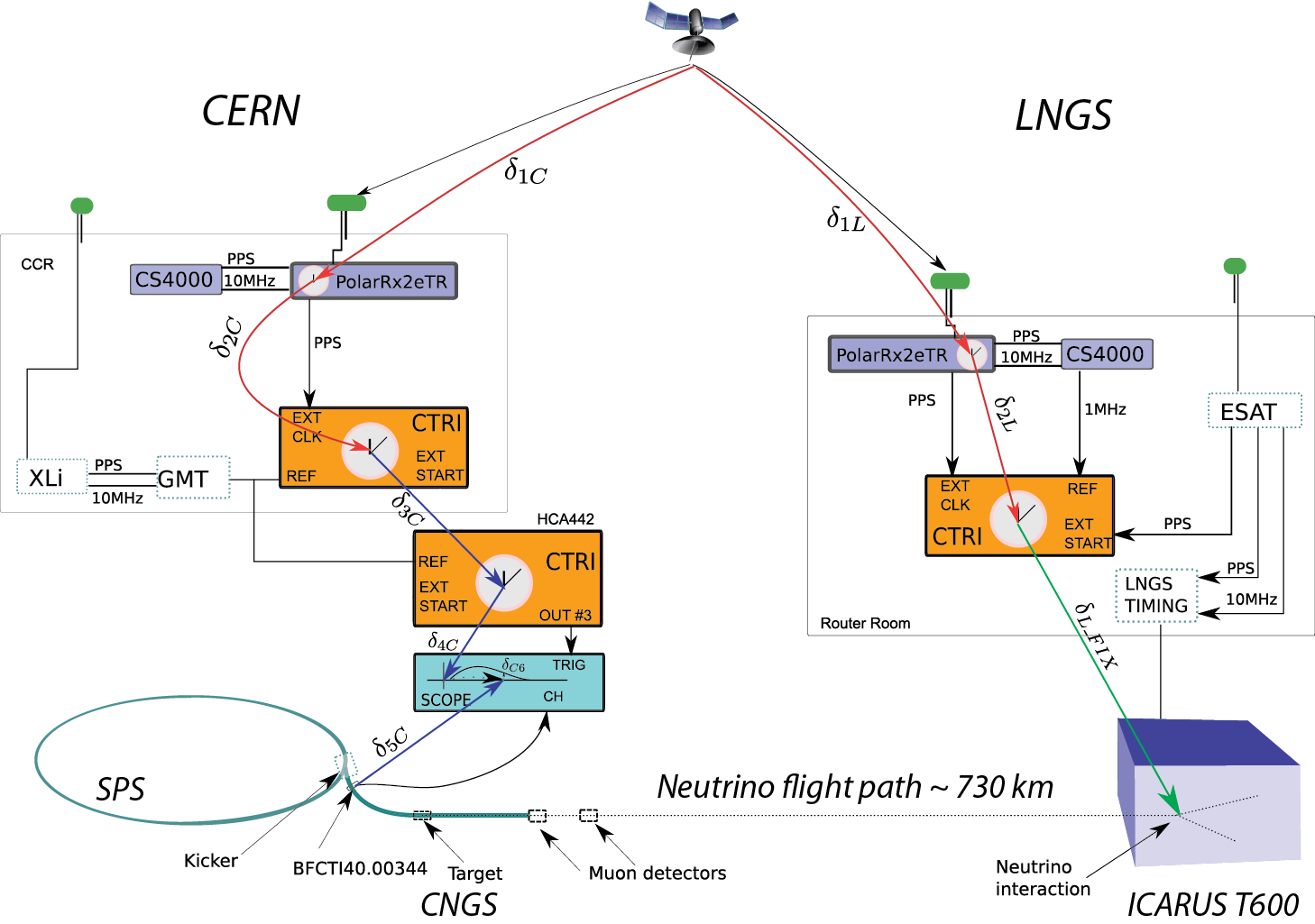}
\caption{{\it General layout of the CERN-LNGS time-link for neutrino time of flight measurement. All timing delays are shown.}}
\label{fig:timelink}
\end{center}
\end{sidewaysfigure}

A detailed description of the CERN and LNGS timing systems and their synchronizations is given in  Ref.~\cite{[01]} and~\cite{[13]}. A schematic picture of the timing systems layout, including all timing delays is shown in Figure~\ref{fig:timelink}. The origin of the neutrino velocity measurement is referred to the Beam Current Transformer (BCT) detector (BFCTI.400344), located $743.391 \pm 0.002$ m upstream of the CNGS neutrino target.

The proton beam time-structure at the BCT is recorded by a 1 GS/s Wave Form Digitiser (ACQIRIS DP110) triggered by the kicker magnet signal. The BFCTI.400344 waveform for every extraction is stored into the CNGS database. Every acquisition is time-tagged with respect to the SPS timing system, associating each neutrino event at LNGS to a precise proton bunch.

The absolute UTC timing signal at LNGS is provided every second (PPS) by a GPS system ESAT 2000 disciplined with a Rubidium oscillator~\cite{[14],[15]}, operating on the surface Laboratory. A copy of this signal is sent underground every ms (PPmS) and used in ICARUS to provide the absolute time-stamp to the recorded events.

In order to improve the $\sim$100 ns time accuracy of the standard GPS at CERN and LNGS, the OPERA Collaboration and CERN have installed, both at CERN and LNGS, two new identical systems composed of a GPS receiver for time-transfer applications Septentrio PolaRx2e operating in Òcommon-viewÓ mode and a Cs atomic clock Symmetricom Cs4000~\cite{[01]}.

The Cs4000 oscillator provides the reference frequency to the PolaRx2e receiver, and a CTRI device logs every second the difference in time between the 1PPS outputs of the standard GPS receiver and of the more precise PolaRx2e, with 0.1 ns resolution. The residual time base fluctuations between the CERN and the LNGS PolaRx2e receivers introduce a systematic error of $\sim$1.0 ns on the ($\delta_{1C}$ + $\delta_{2C}$ - $\delta_{1L}$ - $\delta_{2L}$) calibration factor (Figure~\ref{fig:timelink}). The stability of this calibration was recently confirmed~\cite{[01]}.

The timing signal (PPmS), distributed by the LNGS laboratory, consists of a TTL positive edge (15 ns rise time) sent out every ms and followed, after 200 $\mu$s, by the coding of the absolute time related to the leading edge. This signal is generated in the external laboratory and sent to the underground Hall-B via $\sim$8 km optic fibers. This introduces a delay $\delta L_{\_FIX} =  42036.6 \pm 1.3$~ns, accurately calibrated in December 2011, following a double path procedure very similar to the one devised by the OPERA experiment for their first calibration in 2006~\cite{[01]}.

The method consists in measuring the time difference $\Delta t$ and time sum $\Sigma t$ of the signal propagation along the usual path and an alternative one using a spare fibre.
\begin{itemize}
\item In the first configuration the 1PPS output of the ESAT-2000 GPS receiver was converted into an optical signal, sent underground via the spare fibre and converted back into electrical. The difference in the propagation delays, between this signal and the 1PPmS output of the ESAT-2000 GPS propagated over the standard path, was measured underground taking as a reference the middle height of the rising edge.

\item In the second configuration the 1PPmS output of the ESAT-2000 GPS, at the end of the usual path underground is sent back to the external laboratory, where it is compared with the 1PPS signal, taking as a reference the middle height of the rising edge. 
\end{itemize}
The used optoelectronic chain is kept identical in the two cases by simply swapping the receiver and the transmitter between the two locations. Furthermore the jitter of the phase difference between the 1PPS and the 1PPmS outputs of the ESAT-2000 GPS receiver was checked to be negligible ($\le 0.25$~ns). 

\section{Brief description of ICARUS}
\label{descr}

The ICARUS T600 detector consists of a large cryostat split into two identical, adjacent modules with internal dimensions $3.6 \times 3.9 \times 19.6$ m$^3$ filled with about 760 tons of ultra-pure liquid Argon~\cite{[05],[06],[07]}. Each module houses two TPC's separated by a common cathode that generates an electric field ($E_D$ = 500 V/cm). There are three parallel planes of wires, 3 mm apart with lengths up to 9 m, facing the drift volume 1.5 m long. By appropriate voltage biasing, the first two planes provide signals in a non-destructive way. The third plane collects the ionization charge. There are 53248 channels in total.  Wires are oriented on each plane at a different angle (0$^{\circ}$, $\pm$60$^{\circ}$) with respect to the horizontal direction. Combining the wire coordinate on each plane at a given drift time, a three-dimensional image of the ionizing event is reconstructed. A remarkable resolution of about 1 mm$^3$ is uniformly achieved over the whole detector active volume ($\sim$340 m$^3$ corresponding to 476 t), which allows to precisely determine the neutrino interaction position.

The scintillation light produced by ionizing events in the LAr-TPC is recorded with 74 photomultipliers (PMT) of 8 inches diameter. The PMTÕs are organized in arrays located behind the four wire chambers (1L, 1R, 2L, 2R). The sums of the signals from the PMT arrays (PMT-Sums) are integrated with a fast CANBERRA 2005 preamplifier and used for triggering and to locate the event along the drift direction.  The scintillation decay time from LAr has two distinct components. The fast component (6 ns) records the early detectable photons within 1 ns from the ionization process. The trigger threshold at about 100 photoelectrons detects events with energy deposition as low as few hundred MeV with full efficiency. CNGS neutrinos are recorded within a coincidence between an ``Early Warning Signal" of proton extraction from CERN and the PMT-Sums signals. 

In the present measurement, the PMT readout has been equipped with an additional PMT-DAQ system based on three 2-channel, 8-bit, 1-GHz ACQIRIS AC240 digitizer boards, derived from the DAQ developed for the WArP experiment~\cite{[16]}. At every  CNGS trigger,  the four PMT-Sums waveforms and the absolute time of the LNGS PPmS signal, followed by its related coding, are readout and stored in a local memory buffer of 1400~$\mu$s time depth,  to get both the PMT-Sums and the whole PPmS within the same memory window. 

As the channels on each board are synchronized at the level of a few picoseconds, the time interval between the PMT-Sums and the PPmS signals allows the determination of the absolute time of the scintillation light pulse in the T600 detector within few ns precision. The LNGS time is referred to the half height of the PPmS leading edge (as in the calibration procedure). The time of the PMT-Sums is defined as the onset of the related signal, corresponding to the arrival time of the earliest scintillation photons.

Several additional delay corrections have been included in order to measure the neutrino arrival time. The propagation time of the scintillation light signals from the PMTÕs to the AC240 boards includes the transit time within the PMTÕs (nominal value $\sim$75$\pm$2 ns),  the overall cabling ($\sim$44 m) and the delay through the signal adders and the preamplifier. The values are $\delta_{5L} = 299, 293, 295, 295$ ns for chambers 1L, 1R, 2L, 2R respectively. The whole chain has been measured with the help of LED's immersed in LAr. The associated uncertainty ($\pm$5.5 ns) includes the variations in the PMT transit time due to different biasing voltages and slight differences in cable lengths.

In addition, for each event, the distance of the event from the closest PMT and the position of the interaction vertex along the $\sim$18 m of the detector length have been evaluated. In case of muons from neutrino interaction in upstream rock, the entry point into the active volume is considered as vertex. The ICARUS upstream wall position has been used as reference coordinate for the neutrino time of flight.  The corresponding time corrections ($\delta_{7L}$, $\delta_{6L}$) have been deduced from the event topology in the detector through visual scanning, with an accuracy better than 1 ns. Events in the standard ICARUS DAQ and the new PMT-DAQ have been associated through their common absolute time stamp.

Finally, the position of the ICARUS upstream entry wall (the reference point) has been measured to be 55.7$\pm$0.5~m closer to CERN than the OPERA origin point~\cite{[17]}, resulting in an overall distance of the ICARUS reference point from the CERN beam origin (the position of the BFCTI.400344 intensity monitor) of 731222.3$\pm$0.5~m. Hence, the corresponding neutrino time of flight for $v = c$ is  expected to be $tof_c$ = 2439098$\pm$1.7 ns (including the 2.2 ns contribution due to earth rotation).

\section{Data analysis}
\label{daq}
ICARUS started data taking with the CNGS low intensity bunched beam on October 27$^{th}$, 2011. Unfortunately, due to an initial time shift of 2 ms of the Early Warning Signal information with respect to the actual CERN-SPS proton extraction time, ICARUS T600 recorded neutrino data only from October 31$^{st}$ to November 6$^{th}$, collecting 7 beam-associated events, in rough agreement with the  $2.2 \times 10^{16} $ p.o.t. delivered to CNGS.

\begin{figure}[!htb]
\begin{center}
\includegraphics[width=0.80\textwidth]{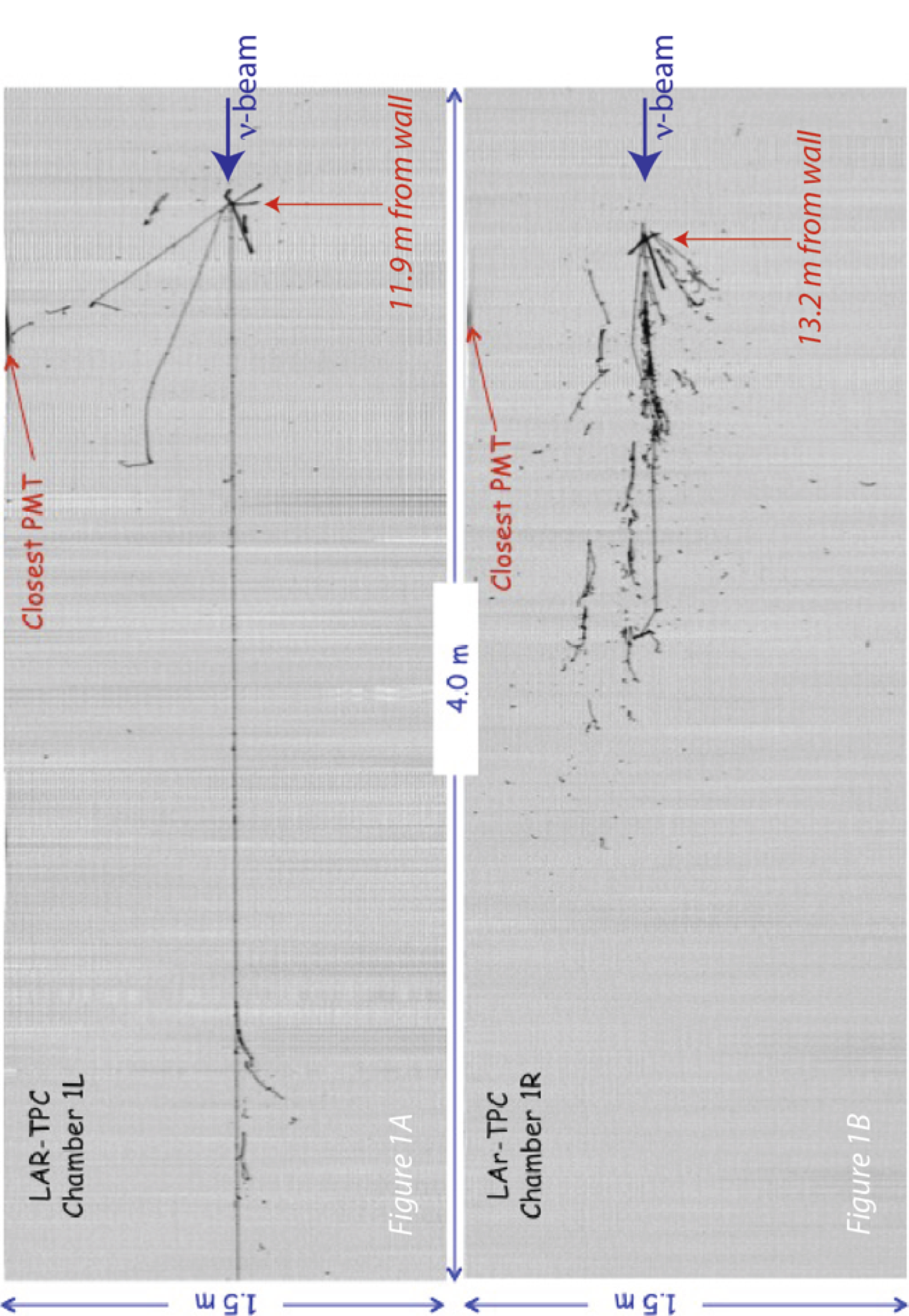}
\caption{{\it (a) CC event in the chamber 1L; (b) NC event in the chamber 1R, as visible in the LAr collection view. The actual distances of the vertex from the upstream wall of the detector are indicated. The  signal of the closest PMT spontaneously induced on the charge collecting wires is also visible.}}
\label{fig:CC_NC}
\end{center}
\end{figure}

The events consist of three neutrino interactions (two CC and one NC) with the vertex contained within the ICARUS active volume and four additional through going muons generated by neutrinos interactions in the upstream rock.  One of the two CC events and the NC one are shown respectively in Figure~\ref{fig:CC_NC}a and \ref{fig:CC_NC}b.

Table~\ref{tab:EVENT} describes in detail the timing of each of the 7 events. The table is divided in the following sections: (1) properties of the ICARUS events; (2) CERN related data; (3) LNGS related data; (4) the time difference between the light and neutrino events. Only variable time corrections are shown.

The proton transit time at the BCT ($T_{START}$) in absolute GPS time base is obtained from the CERN proton extraction time plus fixed and variable corrections: $T_{START}  = T_{CERN} + \delta_{1C} + \delta_{2C} + D_1 + \delta_{3C} + \delta_{4C} + \delta_{5C} + \delta_{6C}$ where $D_1 = -99216$ ns is an overall offset between the CERN and the GPS time-bases, $\delta_{3C} = 10078 \pm 2$ ns, $\delta_{4C} = 27 \pm  1$ ns, $\delta_{5C} = -580 \pm 5$ ns are fixed delays,  $\delta_{1C}$ and $\delta_{2C}$ are the CERN time-link variable corrections,  $\delta_{6C}$ is the BCT signal delay ($\sim$1.3 ns syst. error due to proton beam width).  

The event time in ICARUS ($T_{STOP}$) in absolute GPS time base is obtained from the LNGS distributed absolute time plus fixed and variable corrections: $T_{STOP} = T_{LNGS} + \delta_{1L} + \delta_{2L} - \delta_{3L} + \delta_{4L} -  \delta_{5L} - \delta_{6L} - \delta_{7L} + \delta L_{\_FIX}$, where $\delta L_{\_FIX} = 42037 \pm 1.3$ ns is PPmS propagation fix delay,  $\delta_{1L}$ and $\delta_{2L}$ are the LNGS time-link variable corrections, $\delta_{3L}$ and $\delta_{4L}$ are the PMT-Sum and PPmS signal delays as recorded on the AC240 boards ($\sim$1.5 ns syst. error), $\delta_{5L}$ is the PMT cabling delay (5.5 ns syst. error), $\delta_{6L}$ and $\delta_{7L}$ are the corrections due to vertex position and PMT distance in ICARUS ($\le 1$ ns syst. error).

The $\sim$1.0 ns residual fluctuations on the CERN-LNGS time-base ($\delta_{1C}$ + $\delta_{2C}$ - $\delta_{1L}$ - $\delta_{2L}$) has also to be included as additional systematic error.

The actual neutrino time of flight, $tof_\nu$, is calculated from  $T_{STOP} - T_{START}$ accounting for the additional time related to the nearest proton bunch. The difference $\delta t = tof_c - tof_\nu$ between the expected time of flight based on speed of light ($tof_c = 2439098\pm1.7$ ns) and the actual arrival time of neutrino in the ICARUS detector is shown in the last row.

\begin{figure}[!htb]
\begin{center}
\includegraphics[width=1.00\textwidth]{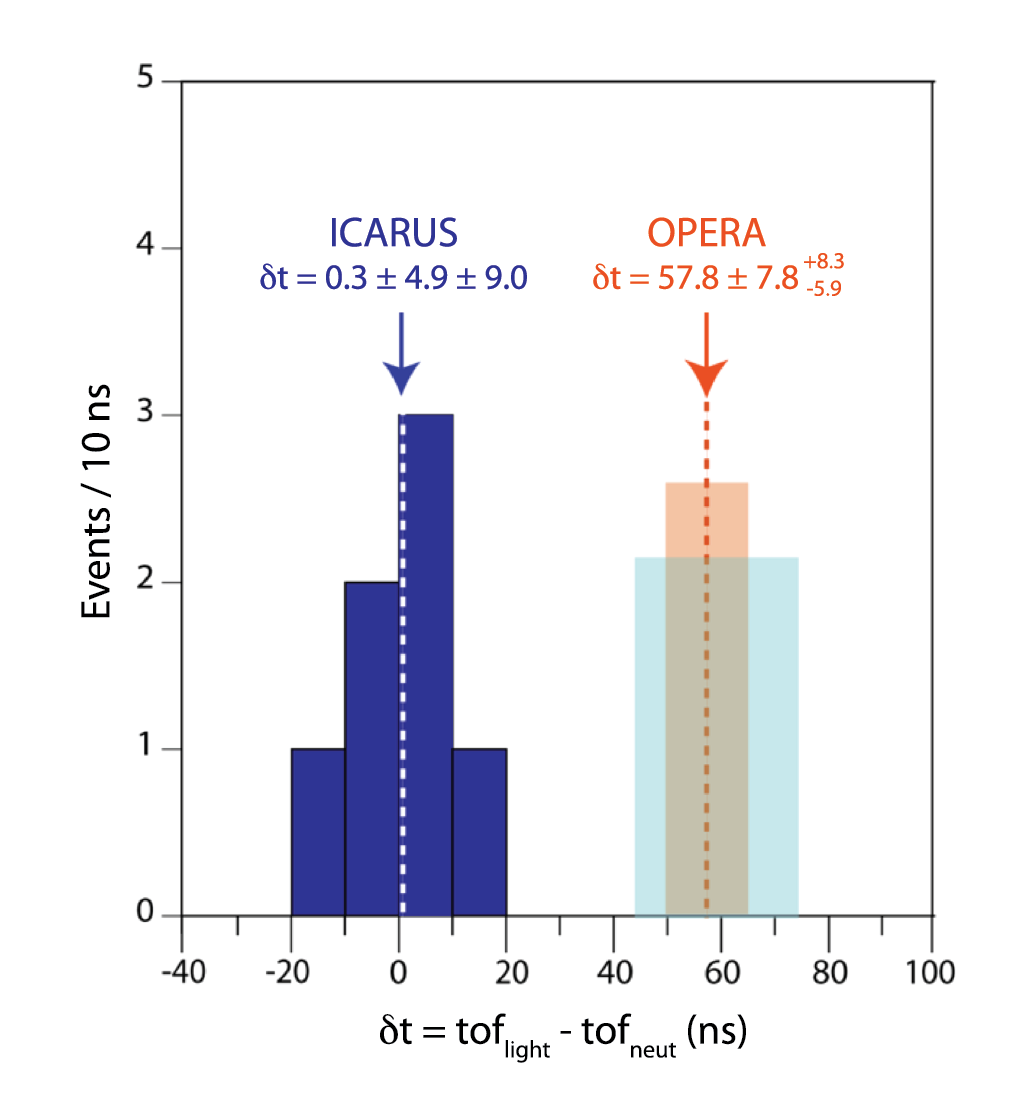}
\caption{{\it Time distribution in ns of the difference $\delta$t  between the neutrino time of flight expectation based on speed of light the actual measurement in the ICARUS LAr-TPC.  Events are compatible with Lorentz invariance, which requires $\delta$t $\le$ 0. The OPERA result is also shown as a comparison.}}
\label{fig:result}
\end{center}
\end{figure}

The average value of $\delta$t for the 7 events is  +0.3 ns with an r.m.s. of 10.5 ns and an estimated systematic error of $\sim$9.0 ns,  obtained from combining in quadrature all previously quoted uncertainties. The statistical error on the average is $4.9$ ns, as estimated from a Student distribution with 6 degrees of freedom (see Figure ~\ref{fig:result}). 

\section{Conclusions}
The expected time of flight difference between the speed of light from CERN to ICARUS and the actual position of the vertex of the LAr-TPC events has been neatly analysed. Based on the seven recorded neutrino events, the result $\delta t = +0.3 \pm 4.9 (stat.) \pm 9.0 (syst.)$~ns is compatible with a neutrino propagation velocity in agreement with the speed of light and incompatible with the result reported by the OPERA Collaboration.

\section*{Acknowledgements}
The ICARUS Collaboration acknowledges the fundamental contribution of INFN to the construction and operation of the experiment. In particular we are indebted to LNGS Laboratory and its Director for the continuous support to the experiment. Moreover we thank the ``Servizi Tecnici'' and ``Servizio Calcolo'' divisions of LNGS for their contribution in preparing the timing facility. We also acknowledge the BOREXINO, LVD and OPERA Collaborations for their valuable cooperation and for the common effort in preparing the future more precise neutrino t.o.f. measurement facility. The Polish groups acknowledge the support of the Ministry of Science and Higher Education in Poland, including project 637/MOB/2011/0 and grant number N N202 064936. Finally we gratefully acknowledge the contribution of CERN, in particular the CNGS staff, for the very successful operation of the neutrino beam facility.

\bibliographystyle{elsarticle-num}


\begin{thebibliography}{00}
\bibitem{[01]} T. Adam et al. [OPERA Collaboration], arXiv:1109.4897.
\bibitem{[02]} P. Adamson et al. [MINOS Collaboration], Phys. Rev. \textbf{D 76} (2007) 072005  [arXiv:0706.0437].
\bibitem{[03]} G. R. Kalbfleisch, N. Baggett, E. C. Fowler and J. Alspector, Phys. Rev. Lett. \textbf{43} (1979) 1361.
\bibitem{[04]} A.G. Cohen, and S.L. Glashow, Phys. Rev. Lett. \textbf{107} (2011) 181803 .
\bibitem{[05]} C. Rubbia, ÒThe Liquid-Argon Time Projection Chamber: A New Concept For Neutrino DetectorÓ, CERN-EP/77-08 (1977). 
\bibitem{[06]} S. Amerio et al. [ICARUS Collaboration], Nucl. Instr. Meth.  \textbf{A527} (2004) 329.
\bibitem{[07]} C. Rubbia et al. [ICARUS Collaboration], JINST  \textbf{6} (2011) P07011.
\bibitem{[08]} M. Antonello et al. [ICARUS Collaboration], arXiv:1110.3763v3, (Submitted to Physics Letters, February 25, 2012).
\bibitem{[09]} R. M. Bionta et al. [IMB Collaboration], Phys. Rev. Lett.  \textbf{58} (1987) 1494.
\bibitem{[10]} E. N. Alekseev, L. N. Alekseeva, I. V. Krivosheina and V. I. Volchenko, Phys. Lett.  \textbf{B 205} (1988) 209.
\bibitem{[11]} K. Hirata et al. [KAMIOKANDE-II Collaboration], Phys. Rev. Lett.  \textbf{58} (1987) 1490.
\bibitem{[12]} J. R. Ellis, N. Harries, A. Meregaglia, A. Rubbia and A. Sakharov, Phys. Rev.  \textbf{D 78}  (2008) 033013 [arXiv:0805.0253].
\bibitem{[13]} P. Alvarez Sanchez and J. Serrano, CERN BE-CO-HT Internal Note, (Oct. 16, 2011).
\bibitem{[14]} Master clock GPS 2000, http://www.esat.it/EN/default.htm, M. Ambrosio, et al., Phys. Rev. \textbf{D 62} (2000)  052003.
\bibitem{[15]} M. Ambrosio et al., The MACRO detector at Gran Sasso, Nucl. Instr. Meth. \textbf{A 486}  (2002) 663.
\bibitem{[16]} R. Acciarri et al., Journal of Phys. Conf. Series \textbf{308} (2011) 012005 and references therein. 
\bibitem{[17]} G. Brunetti, "Neutrino velocity measurement with the OPERA experiment in the CNGS beamÓ, PhD ThesisÊ(2011);  G. Colosimo et al., ÒDetermination of the CNGS global geodesyÓ, OPERA public note 132 (2011). 
\end{thebibliography}







\begin{sidewaystable}[!htb]
\hspace{-1.0cm}
\begin{tabular}{>{\scriptsize}l>{\scriptsize}c>{\scriptsize}c>{\scriptsize}c>{\scriptsize}c>{\scriptsize}c>{\scriptsize}c>{\scriptsize}c}
\hline\noalign{\smallskip}
  Event $\#$       &    1         &     2     &     3           &     4        &     5     &     6     &       7     \\
\noalign{\smallskip}\hline\noalign{\smallskip}
{\sl \hspace{1cm} General properties of the event}  &           &             &            &          &          \\
  Date                            & 31-Oct       &    1-Nov     &     2-Nov    &     2-Nov    &     2-Nov &    2-Nov  &    4-Nov    \\
  Time (UTC+1)                    & 08:41:22.554 & 03:57:00.954 & 09:11:56.154 & 02:49:16.556 &  11:18:51.356 & 16:53:41.756 & 23:31:27.356\\
  Run number                      &  10949       &     10949    &   10949      &     10950    &     10951  &   10951  &    10956    \\ 
  Event number                    &    338       &     1247     &    1885      &    1053      &       91   &    802   &      773    \\
  Type of event                   &   Mu-rock    &   Nu-NC      &  Nu-CC       &    Mu-rock   &    Nu-CC   &   Mu-rock&    Mu-rock  \\
  TPC Chamber (PMT array)         &  1R(1L)      &   1R(1L)     &  1L(1L)      &     2L(2R)   &    1L(1L)  &   2L(2R) &  2L(2R)     \\
{\sl \hspace{1.0cm} CERN related data} &         &              &              &              &            &          &             \\
 CERN extraction time, $T_{CERN}$ (CERN time base, ns) & 133004  & 113399 & 114172  &  131046      &   116852   &  117040  & 118340      \\
 $1^{st}$ CERN time link correction, $\delta_{1C}$ (ns) & 2443   &  2457  & 2464 &  2487        &    2500    &  2507    & 2558        \\
 $2^{nd}$ CERN time link correction, $\delta_{2C}$ (ns) & 86603  &  86590 & 86585& 86564        &   86556    &  86537   & 86500       \\
 BCT delay (first bunch), $\delta_{6C}$ (ns)           & 6047   &  6044  & 6049 & 6055         &   6050     &   6041   & 6046        \\
 Proton transit time at BCT, $T_{START}$ (GPS time base, ns)  & 138406 & 118799 & 119579 & 136461& 122267   &122434    & 123753      \\
{\sl \hspace{1.0cm} LNGS related data}   &              &             &               &              &            &          &       \\
LNGS distributed time, $T_{LNGS}$ (LNGS time base, ns) & 3000000 & 3000000 & 3000000 & 3000000 &   3000000 & 3000000  & 3000000     \\
$1^{st}$ LNGS time link correction, $\delta_{1L}$ (ns)& -7316    & -7317 & -7321 & -7318        & -7313      & -7312    & -7323       \\
$2^{nd}$ LNGS time link correction, $\delta_{2L}$ (ns)& 7035     & 7030  &  7062 &  7029        &  7064      &  7060    &  7041       \\
LNGS time on AC240, $\delta_{3L}$ (ns)            &   471602  & 491167   &  489889              & 474031  & 487231  & 486570  & 486296\\
PMT time on AC240, $\delta_{4L}$ (ns) & $8196^{(1)}$&      8193  &          8194 &         8190 &      8185  &   8204   &    8212     \\
PMT cable delay, $\delta_{5L}$ (ns)   &   299     &      299     &   299         &         295  &       299  &    295   &    295      \\
Vertex position correction, $\delta_{6L}$ (ns)&  0    &     40   &          44   &       16     &       28   &    5     &     1       \\
PMT position correction, $\delta_{7L}$ (ns)   &  12   &     11   &          5    &       13     &        4   &    13    &     13      \\
Event time in ICARUS, $T_{STOP}$ (GPS time base, ns)& 2578030  & 2558421 &  2559730 & 2575578 & 2562406    & 2563101  & 2563357     \\
{\sl \hspace{1.0cm} Neutrino time-of-flight calculation}       &             &          &           &         &          &     &     \\ 
$T_{STOP}$-$T_{START}$ &2439632 & 2439622 & 2440151 & 2439117 & 2440139 & 2440667 & 2439604 \\
 Nearest proton bunch             &     2       &      2       &     3    &     1     &     3   &    4     &   2      \\
Bunch related additional time (ns)&    524      &     524      &    1048  &     0     &   1048  &   1572   &  524      \\
Neutrino time of flight ($tof_{\nu}$, ns) &   2439104    &  2439098    &  2439103   &  2439117  & 2439091  & 2439095 & 2439080 \\
Time difference w.r.t. expectation ($\delta t = tof_c - tof_{\nu}$, ns)  &  -6     &   0    &   -5    &  -19  &   +7   &  +3   & +18  \\
 \noalign{\smallskip}\hline
 \end{tabular}
 \caption{{\it Events collected with CNGS bunched beam. The actual neutrino time of flight is calculated from $T_{STOP} - T_{START}$ accounting
for the additional time related to the nearest proton bunch. The difference $\delta t = tof_c - tof_{\nu}$ with respect to the expected
time of flight based on the speed of light ($tof_c = 2439098\pm1.7$ ns) is shown in the last row. For details we refer to the text.
$^{(1)}$The value for event  $\# 1$ has a larger uncertainty of about 4 ns (about the double w.r.t. the other events), due to the smaller PMT  pulse height.}}
 \label{tab:EVENT}
 \end{sidewaystable}

\end{document}